\documentclass[aps,prl,twocolumn,showpacs,superscriptaddress,citeautoscript]{revtex4-1} %preprint
\usepackage{graphicx}
\usepackage{amsmath,amssymb,amsfonts}
\usepackage{textcomp}
\usepackage{hyperref}
\usepackage{gensymb}
\usepackage{verbatim}
\usepackage{amsmath}
\hypersetup{breaklinks=true,colorlinks=true,urlcolor=black}
\usepackage{color}

\newcommand{\YbFeZn} {YbFe$_2$Zn$_{20}$}
\newcommand{\Tc} {$T_\textrm{M}$}
\newcommand{\Tmax} {$T_\textrm{max}$}
\newcommand{\Tmaxa} {$T_\textrm{max1}$}
\newcommand{\Tmaxb} {$T_\textrm{max2}$}

\newcommand{\pc} {$p_c$}

\begin{document}

	\title{Collapse of Kondo state and ferromagnetic quantum phase transition in YbFe$_2$Zn$_{20}$}
	\author{Udhara~S. \surname{Kaluarachchi}}  
	\affiliation{Ames Laboratory, US Department of Energy, Iowa State University, Ames, Iowa 50011, USA}
	\affiliation{Department of Physics and Astronomy, Iowa State University, Ames, Iowa 50011, U.S.A.} 
	\author{Li \surname{Xiang}}
	\affiliation{Ames Laboratory, US Department of Energy, Iowa State University, Ames, Iowa 50011, USA}
	\affiliation{Department of Physics and Astronomy, Iowa State University, Ames, Iowa 50011, U.S.A.} 
	\author{Jianjun \surname{Ying}}  
	\affiliation{Geophysical Laboratory, Carnegie Institution of Washington, Washington, DC 20015, USA}
	\affiliation{HPCAT, Geophysical Laboratory, Carnegie Institution of Washington, Argonne, Illinois 60439, USA}
	\author{Tai \surname{Kong}}  
	\affiliation{Ames Laboratory, US Department of Energy, Iowa State University, Ames, Iowa 50011, USA}
	\affiliation{Department of Physics and Astronomy, Iowa State University, Ames, Iowa 50011, U.S.A.} 
	\author{Viktor \surname{Struzhkin}}  
	\affiliation{Geophysical Laboratory, Carnegie Institution of Washington, Washington, DC 20015, USA}
	\author{Alexander  \surname{Gavriliuk }}  
	\affiliation{FSRC “Crystallography and Photonics” of Russian Academy of Sciences, Moscow 119333, Russia}
	\affiliation{Institute for Nuclear Research, Russian Academy of Sciences, Moscow, Troitsk, 108840, Russia}
	\affiliation{REC «Functional Nanomaterials», Immanuel Kant Baltic Federal University, 236041 Kaliningrad, Russia}
	\author{Sergey~L. \surname{Bud'ko}}  
	\affiliation{Ames Laboratory, US Department of Energy, Iowa State University, Ames, Iowa 50011, USA}
	\affiliation{Department of Physics and Astronomy, Iowa State University, Ames, Iowa 50011, U.S.A.}
	\author{Paul~C. \surname{Canfield}}
	\affiliation{Ames Laboratory, US Department of Energy, Iowa State University, Ames, Iowa 50011, USA}
	\affiliation{Department of Physics and Astronomy, Iowa State University, Ames, Iowa 50011, U.S.A.}

	\date{\today}
	%\email[]{afasdfasd}

	\begin{abstract}
We present the electrical resistivity data under application of pressures up to $\sim$\,26\,GPa and down to 50\,mK temperatures on \YbFeZn. We find a pressure induced magnetic phase transition with an onset at \pc\,=\,18.2$\pm$\,0.8\,GPa. At ambient pressure, YbFe$_2$Zn$_{20}$ manifests a heavy fermion, nonmagnetic ground state and the Fermi liquid behavior at low temperatures. As pressure is increased, the power law exponent in resistivity, $n$, deviates significantly from Fermi liquid behavior and tends to saturate with $n$\,=\,1 near \pc. A pronounced resistivity maximum, \Tmax{}, which scales with Kondo temperature is observed. \Tmax{} decreases with increasing pressure and flattened out near \pc{} indicating the suppression of Kondo exchange interaction. For $p$\,$>$\,\pc{}, \Tmax{} shows a sudden upward shift, most likely becoming associated with crystal electric field scattering. Application of magnetic field for $p$\,$>$\,\pc{} broadens the transition and shifts it toward the higher temperature, which is a typical behavior of the ferromagnetic transition. The magnetic transition appears to abruptly develop above \pc{}, suggesting probable first-order (with changing pressure) nature of the transition; once stabilized, the ordering temperature does not depend on pressure up to $\sim$ 26 GPa. Taken as a whole, these data suggest that YbFe$_2$Zn$_{20}$ has a quantum phase transition at \pc{} = 18.2 GPa associated with the avoided quantum criticality in metallic ferromagnets.

	\end{abstract}
\maketitle

	\section{Introduction}

	Among the rare-earth-based intermetallic compounds, Ce and Yb -based materials have attracted much attention due to their peculiar properties\,\cite{Movshovich1994PRL,Cornelius1995PRB,Grosche1996,Mathur1998Nature,AlamiYadri1998,Winkelmann1999PRB,Knebel2001PRB,Stewart2001RMP,Budko2004PRB,Coleman2007,Nakatsuji2008NatPhy,Flouquet2009arxIv,Si2010,Steppke2013Science}. The properties of these compounds are usually dominated by two characteristic energy scales: Ruderman-Kittel-Kasuya-Yosida (RKKY)\,\cite{Ruderman1954PhyRev,Kasuya1956,Yosida1957PhyRev} and Kondo\,\cite{Kondo1964,Hewson1993} interaction energies. The exchange interaction, $J$, which determines the interaction energy between local moments and conduction electron, $T_\textrm{K}$\,$\propto$\,e$^{-1/J}$, is also responsible for the coupling between local moments through the RKKY interaction, $T_\textrm{RKKY}$\,$\propto$\,$J^2$. Hence, the ground state of these compounds is determined by the competition between these two energy scales and often described by the Doniach phase diagram\,\cite{Doniach1977}. When $T_\textrm{K}$\,$\gg$\,$T_\textrm{RKKY}$ the ground state in nonmagnetic and when $T_\textrm{K}$\,$\ll$\,$T_\textrm{RKKY}$, magnetic order can be established. The most interesting situation occurs when the two energy scales are comparable and the system can be tuned through a $T$\,=\,0\,K magnetic instability at a quantum phase transition (QPT). Application of external pressure is one of the ways to tune the $J$\,\cite{Schilling1979,Jackson2005PRB}, bringing the material to a QPT or QCP (quantum critical point).

	Often Yb is considered as a "hole" equivalent of Ce. In contrast to Ce compounds, where magnetic ordering is suppressed by pressure, in Yb systems, increasing pressure can tune the system from a nonmagnetic state to a magnetic one\,\cite{Stewart2001RMP,Flouquet2009arxIv}. There are only a few examples of the pressure induced, nonmagnetic-to-magnetic phase transitions in Yb compounds\,\cite{Winkelmann1998PRL,Nakano2004,Winkelmann1999PRB,AlamiYadri1998,Schoeppner1986,Saiga2008JPSJ,Cornelius1995PRB, Knebel2001,Yuan2006PRB} and, so far, superconductivity has been reported in only two materials\,\cite{Nakatsuji2008NatPhy, Schuberth2016}.
	
%with a Yb$^{3+}$ state
	The Yb$T_2$Zn$_{20}$ ($T$\,=\,Fe, Ru, Os, Co, Rh, Ir) series is a Yb-based heavy fermion system\,\cite{Torikachvili2007PNAS,Jia2008PRB,Canfield2008,Jia2009PRB,Mun2012PRBb} which belongs to the  $R$$T_2$Zn$_{20}$ family\,\cite{Nasch1997}. For all six members, at high temperature, the magnetic susceptibility measurements show Curie-Weiss behavior with the effective moment close to the Yb$^{3+}$\,\cite{Torikachvili2007PNAS,Jia2008PRB}. In the resistivity measurements, there are no signs for the magnetic ordering down to 20\,mK\,\cite{Torikachvili2007PNAS,Jia2008PRB}. Among this Yb$T_2$Zn$_{20}$ series, YbCo$_2$Zn$_{20}$ has the lowest $T_\textrm{K}$ and the largest Sommerfeld coefficient of the six members\,\cite{Torikachvili2007PNAS,Jia2008PRB}. By combining the Doniach model with this small $T_\textrm{K}$ and large Sommerfeld coefficient, one can assume that YbCo$_2$Zn$_{20}$ is close to a possible magnetic QCP. With this idea, Saiga $et\,al.$\,\cite{Saiga2008JPSJ} performed a high pressure resistivity measurement on YbCo$_2$Zn$_{20}$ and observed a pressure induced QCP at a critical pressure $\sim$\,1\,GPa and antiferromagnetic (AFM) ordering at higher pressures. Apart from this, a field induced ordered phase has been observed at the ambient pressure in YbCo$_2$Zn$_{20}$, possibly due to the crystal electric field level crossing\,\cite{Shimura2011JPSJ,Takeuchi2011JPSJ,Shimura2012JPSJ,Taga2012JPSJ,Takeuchi2011JPSJb, Honda2014JPSJ}. A pressure induced QCP has also been estimated for YbIr$_2$Zn$_{20}$ ($\approx$\,5.2\,GPa) and YbRh$_2$Zn$_{20}$ ($\approx$\,5.2\,GPa), however, for YbRh$_2$Zn$_{20}$ no pressure induced magnetic transitions have been observed so far\,\cite{Matsubayashi2009,Honda2010JPSJ,Honda2012JPSJ,Honda2013JPSJ}.
	
	Several years ago, high pressure resistivity measurements were performed up to 8.23\,GPa for YbFe$_2$Zn$_{20}$\,\cite{Kim2013PRB88}. Increasing pressure drives $T_\textrm{K}$ to lower values and enhanced the  $A$-coefficient ($\Delta\rho(T)\propto AT^2$); a  QCP of $\sim$ 10 GPa was inferred\,\cite{Kim2013PRB88}. In this work, by employing a diamond anvil cell in a dilution refrigerator, we extend the pressure range up to $\sim$\,26\,GPa and lower the base temperature to 50\,mK. As a result we find a clear feature in resistivity that we identify as a magnetic phase transition in \YbFeZn{} for $p$\,$>$\,18.2\,GPa. The transition temperature is about 1\,K and does not change with further increase of pressures up to 26\,GPa. We tentatively identify the transition as ferromagnetic in nature and associate the step-like feature in $T_\text C(p)$ with an avoided quantum criticality QPT.

	 %meta magnetic YbIr$_2$Zn$_{20}$ and YbRh$_2$Zn$_{20}$\cite{Yoshiuchi2009JPSJ,Takeuchi2010JPSJ}
	
	% Although similar behaviors in the resistivity measurements were observed
%	pressure-induced magnetism in nonmagnetic Yb compounds,
%	e.g., Yb2Ni2Al,7\cite{Winkelmann1998PRL} YbCu2Si2,8\cite{Winkelmann1999PRB} and YbPd2Si2.9\cite{Nakano2004} YbCuAl\cite{Schoeppner1986} YbNiSn\cite{Cornelius1995PRB}

%YbCuAl -intermediate valnce
%YbCu2Si2 -intermediate valnce
%YbAgGe- Yb3+  ordering near 0.6 K
%YbPtIn-Yb3+ ordering near 2.1 K
%YbNi2B2C- Yatskar1996

%   \cite{Aronson2010} \,\cite{Mun2010PHD}
	
%	The physical properties of these compounds can be
%	compared to the Ce-compounds using the electron-hole
%	analogy.  In this picture the missing 4f electron in the
%	4f 13 configuration of the Yb3+-ion can be interpreted as
%	the presence of a 4f-hole, analogously to the 4f-electron in
%	the Ce3+-ion.

%This Doniach phase diagram is well followed by many Ce-based\,\cite{Iglesias1997PRB,Graf1997PRL,Nakashima2004,Ragel2007,Pfleiderer2009RMP,Weng2016,Hayashi2016JPSJ,Cornelius1994PRB,Cornelius1997PRB}  as well as Yb compounds\,\cite{AlamiYadri1998,Colombier2009PRBYbCu2Si2,Flouquet2009arxIv,Knebel2006JPSJ,Cornelius1995PRB,Winkelmann1999PRB,Saiga2008JPSJ}.

	\section{Experimental Methods}
	Single crystals used for this study were grown using a high-temperature solution growth technique\,\cite{CANFIELD1992,Canfield2001} with the help of frit-disc crucible set\,\cite{Canfield2016}. More details about the crystal growth can be found in Refs.\,\onlinecite{Jia2007Nat,Torikachvili2007PNAS,Kong2017PRB}. Temperature and field dependent resistivity measurements were carried out using a Quantum Design Physical Property Measurement System from 1.8\,K to 300\,K. A dilution refrigerator option was utilized to perform measurements down to 50\,mK. The resistivity was measured using the van der Pauw method\,\cite{Pauw1958,Ramadan1994} with ac-current ($I$\,=\,0.005\,mA, $f$\,=\,18.3 and 21.3\,Hz) parallel to the [111] plane and a magnetic field was applied perpendicular to the current plane. A miniature diamond anvil cell\,\cite{Gavriliuk2009RSI}, with 300\,$\mu$m culets, was used to generate the pressure for the resistivity measurement and KCl powder was used as a pressure transmitting medium. The temperature gradient between the dilution refrigerator thermometer and sensor positioned on miniature diamond anvil cell close to the anvils was evaluated in a separate experiment and was found negligible with the protocol of the measurement used in this work. Single crystals with a typical dimension of 80$\times$80$\times$20\,$\mu$m$^3$ were loaded into the sample chamber with an inner diameter of 130\,$\mu$m made out of a cubic BN gasket. Pressure was applied at room temperature and ruby fluorescence, at 300\,K was used to determine the pressure\cite{Piermarini1975JAP}.

	\section{Results and Discussion}

%	\begin{figure}
%		\begin{center}
%			\includegraphics[width=8.5cm]{ALL_rho_v1}
%		\end{center}
%		\caption{\label{Rho_T}(Color online) Temperature dependence of the resistivity of three different samples of \YbFeZn: (a) sample 1, (b) sample 2 and (c) sample 3.} 
%	\end{figure}

	\begin{figure}
		\begin{center}
			\includegraphics[width=8.5cm]{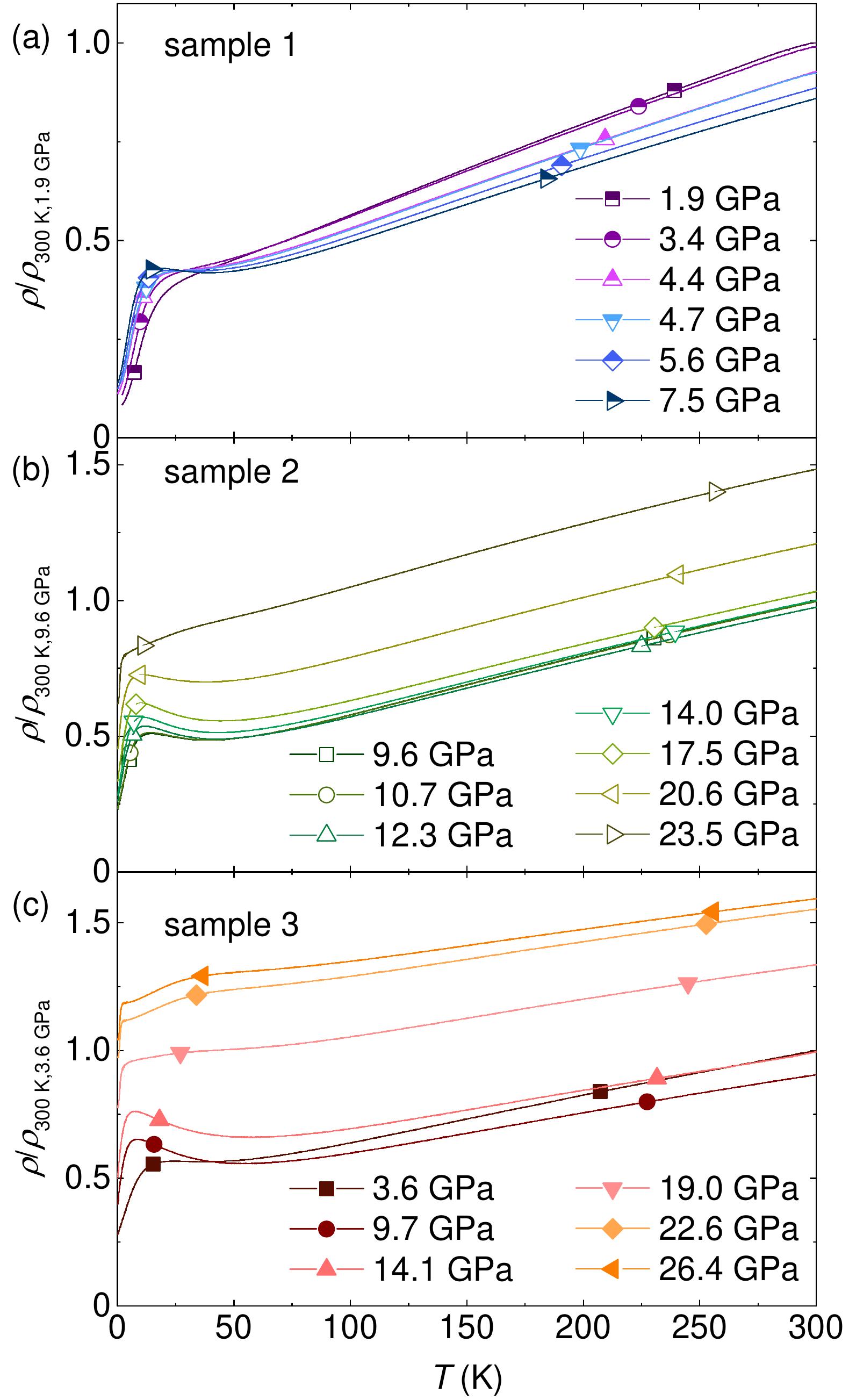}
		\end{center}
		\caption{\label{Rho_T}(Color online) Temperature dependence of the normalized resistivity of three different samples of \YbFeZn: (a) sample 1, (b) sample 2 and (c) sample 3.  Data have been normalized to the lowest pressure 300\,K resistivity value of each sample. Un-normalized curves are shown in appendix (Fig.\,\ref{Rho_T_UN}).} 
	\end{figure}

	\begin{figure}
		\begin{center}
			\includegraphics[width=8.5cm]{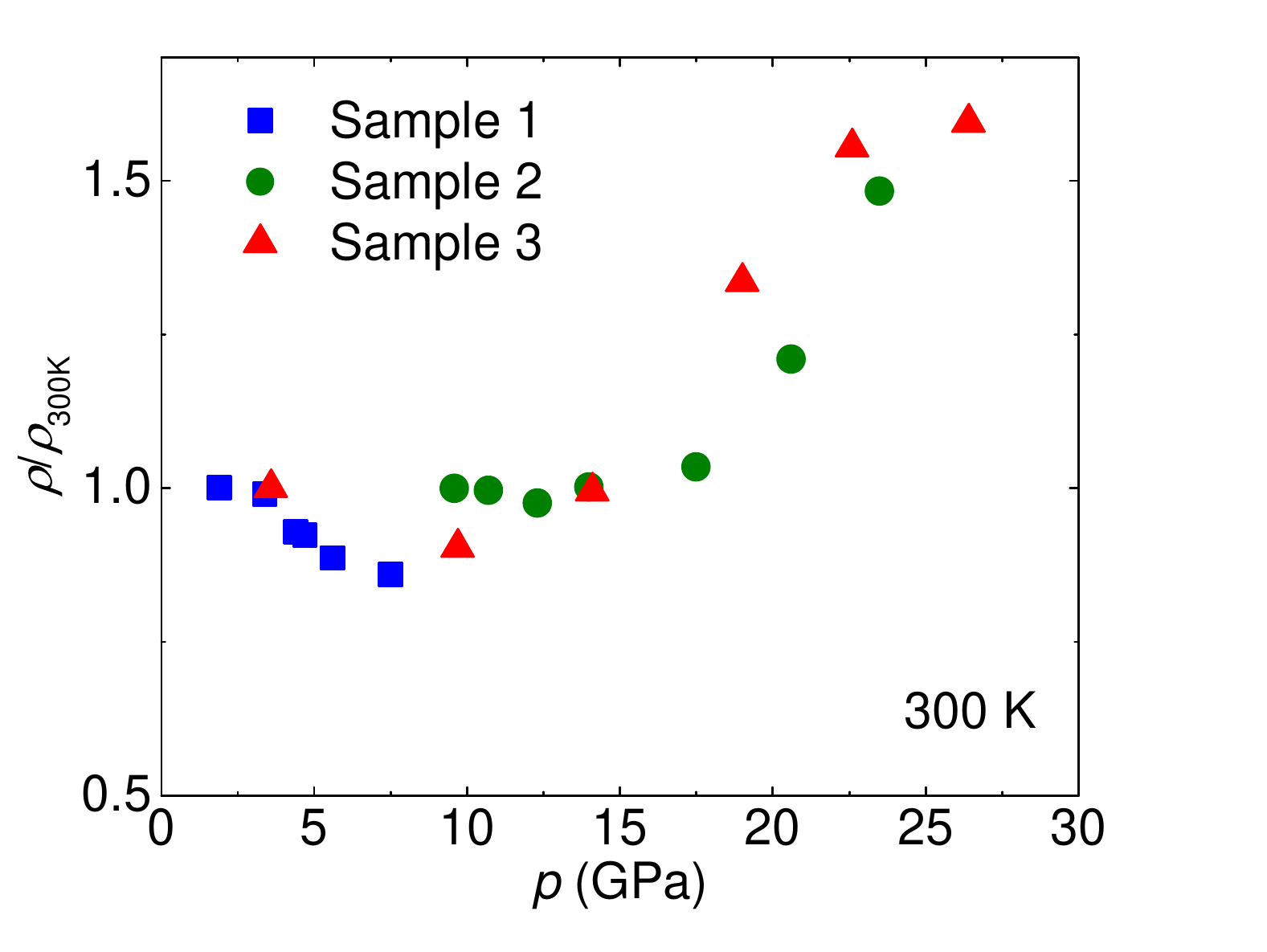}
		\end{center}
		\caption{\label{R300K}(Color online) Pressure dependence of the normalized resistivity at 300\,K. For each samples, $\rho$ is normalized by the lowest pressure 300\,K resistivity value of each sample. Un-normalized curves are shown in appendix (Fig.\,\ref{RRR} (c)).}
	\end{figure}

	\begin{figure}
		\begin{center}
			\includegraphics[width=8.5cm]{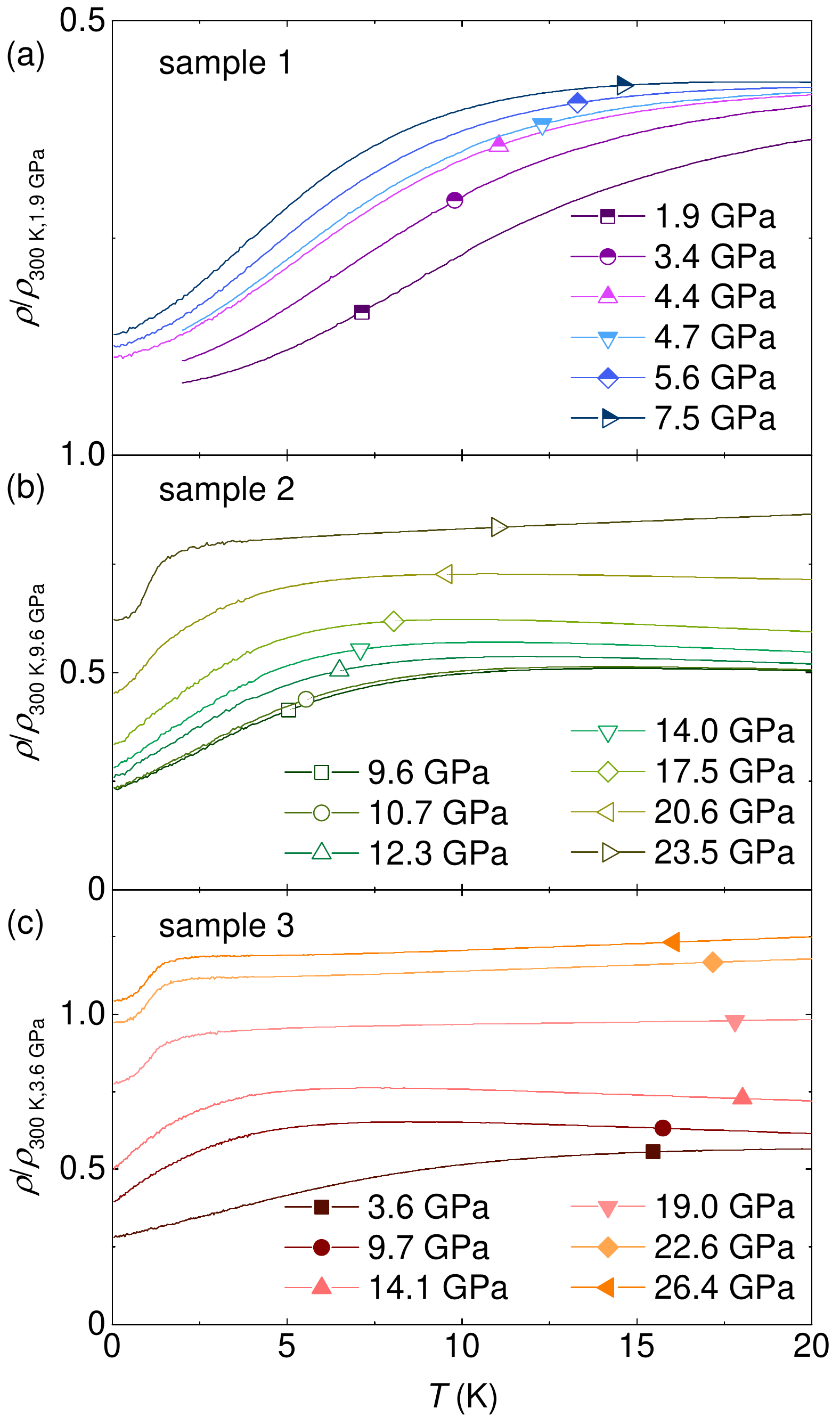}
		\end{center}
		\caption{\label{Rho_low_T}(Color online) Blowup of the low temperature resistivity as in Fig.\,\ref{Rho_T}. Resistivity was measured down to 1.8 K for the first three pressures of sample 1, all the others have been measured down to 50 mK.}
	\end{figure}

%	\begin{figure}
%		\begin{center}
%			\includegraphics[width=8.5cm]{RT_DAC_3_4_compare_linear_v2}
%		\end{center}
%		\caption{\label{Rho_T_comp}(Color online) Comparison of the resistivity data between sample 2 (left axis) and sample 3 (right axis).} 
%	\end{figure}
	Temperature dependent resistivity measurements  on three different samples of \YbFeZn{} under pressures up to 26.4\,GPa are shown in  Fig. \ref{Rho_T}. For each pressure, resistivity values are normalized to the lowest pressure, 300\,K resistivity value of each sample. For sample 1, when increasing the pressure, the 300\,K resistivity, $\rho_{300K}$, is monotonically suppressed, which is similar to Ref.\,\onlinecite{Kim2013PRB88}. For sample 2 and 3, as indicated in Figs. \ref{Rho_T} (b) and (c), $\rho_{300K}$ shows a non-monotonic dependence on pressure when higher pressure values are achieved. However, for $p \lesssim$ 18 GPa, $\rho_{300K}$ shows relatively small variation with pressure, while for $p \gtrsim$ 18 GPa, $\rho_{300K}$ systematically increases with pressure. Fig. \ref{R300K} presents the pressure evolution of the normalized $\rho_{300K}$ with pressure. As indicated in the figure, $\rho_{300K}$ stays relatively flat for $p \lesssim$ 18 GPa and continuously increases for $p \gtrsim$ 18 GPa.  Fig. \ref{Rho_low_T} presents a low temperature expanded view of the data presented in Fig.\,\ref{Rho_T}. In addition to the increased $\rho_{300K}$ for $p$\,$\gtrsim$\,18\,GPa there is also the clear onset of a relatively sharp, low temperature feature (Fig. \ref{Rho_low_T}) for $p$\,$\gtrsim$\,20\,GPa. Whereas these are qualitative changes we will now examine these data quantitatively.

	For all measured pressures, for $T$\,$>$\,50\,K, the resistivity data show a nearly linear temperature dependence (Fig.\,\ref{Rho_T}). It is worth noting that the high-temperature slope (250\,K$<$\,$T$\,$<$\,300\,K) of the resistivity decreases with increasing pressure up to about 10\,GPa and then remains constant for higher pressures.  Below 50\,K, there is a broad shoulder in the resistivity data for $p$\,$<$\,3.4\,GPa that changes into a broad maximum (\Tmax{}) with pressure increasing above 3.6 GPa. The value of \Tmax{} usually scales with the Kondo temperature, $T_\textrm{K}$. It moves to lower temperatures with increasing pressure up to about 20\,GPa and  then shows a sudden increment for $p$\,$>$\,20\,GPa. The behavior of pressure dependence of the \Tmax{} (for $p< 9.6$ GPa) is consistent  with the previous work\,\cite{Kim2013PRB88}. 

	The total resistivity of the \YbFeZn{} can be expressed as a combination of normal metallic behavior and a magnetic contribution. As mentioned above, the high-temperature resistivity shows a nearly linear temperature dependence, indicating that phonon scattering is dominant in the high temperature range. Normal metallic behavior can be approximated by considering the temperature dependent resistivity of non-magnetic LuFe$_2$Zn$_{20}$. Therefore, the magnetic contribution to the resistivity of \YbFeZn{} can be estimated by subtracting the LuFe$_2$Zn$_{20}$ resistivity data from YbFe$_2$Zn$_{20}$ data. Since the residual resistivity values of our samples show non-monotonic increments, first we have to subtract their residual resistivity values for each data set and then normalize the high temperature ($T$\,$>$\,275\,K) slope of the resistivity to that of LuFe$_2$Zn$_{20}$. This can be written as

	\begin{equation}
	\rho_{mag}(T)=(\rho_\textrm{Yb}-\rho_\textrm{Yb,0})\frac{\frac{\textrm{d}\rho_\textrm{Lu,R}}{\textrm{d}T}}{\frac{\textrm{d}\rho_\textrm{Yb,R}}{\textrm{d}T}}\bigg\rvert_{275 K}-(\rho_\textrm{Lu}-\rho_\textrm{Lu,0})
	\label{Eq_rhomag}
	\end{equation} 

\begin{figure}[htb]
	\begin{center}
		\includegraphics[width=8.5cm]{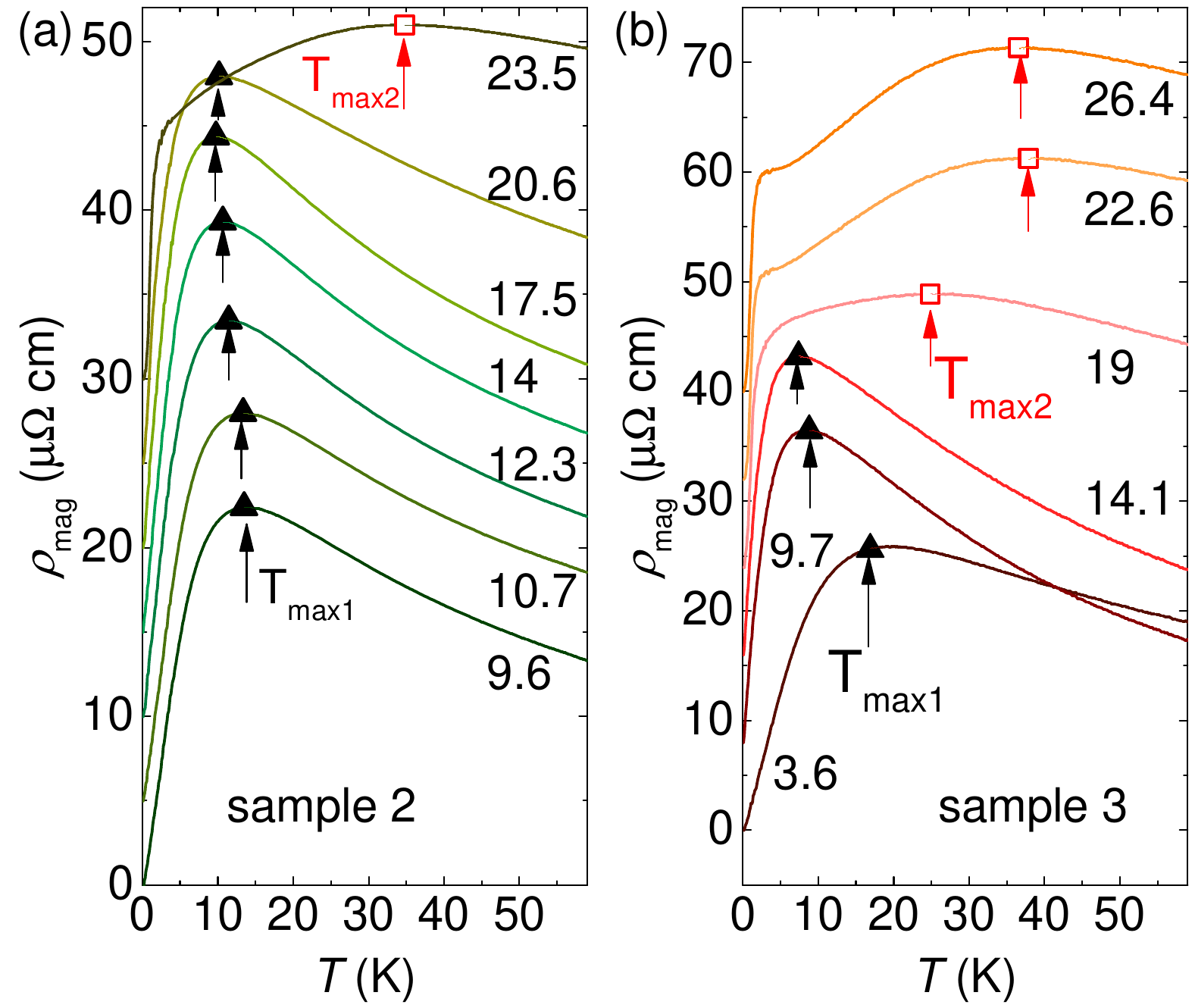}
	\end{center}
	\caption{\label{Rho_mag}(Color online) Temperature dependent $\rho_{mag}$ for (a) sample 2 and (b) sample 3. $\rho_{mag}$ is obtained from Eq.\,\ref{Eq_rhomag} (see text below). The solid black triangles$\backslash$arrows and open red squares$\backslash$arrows indicate \Tmaxa{} and \Tmaxb{} respectively. Data are offset for clarity.} 
\end{figure}

	\noindent Similar analysis has been used to determine the $\rho_{mag}(T)$ in Ref.\,\onlinecite{Jia2009PRB}. $\rho_{mag}(T)$ data for samples 2 and 3 are shown in  Figs.\,\ref{Rho_mag}\,(a) and (b) respectively. For both samples, $T_\text{max1}$ decreases with increasing pressure up to about 20\,GPa  and then shows a sudden change (jump up in temperature) for higher pressures. $T_\text{max2}$ is the temperature corresponding to this higher pressure, broad, maximum in the $\rho_{mag}(T)$ low temperature data. The solid black triangles and open red squares indicate the \Tmaxa{} and \Tmaxb{} respectively.
%\cite{Winkelmann1998PRL,Winkelmann1999PRB,Knebel2001,Yuan2006PRB,Yadri1998,Plessel2003PRB,}

\begin{figure}
	\begin{center}
		\includegraphics[width=8.5cm]{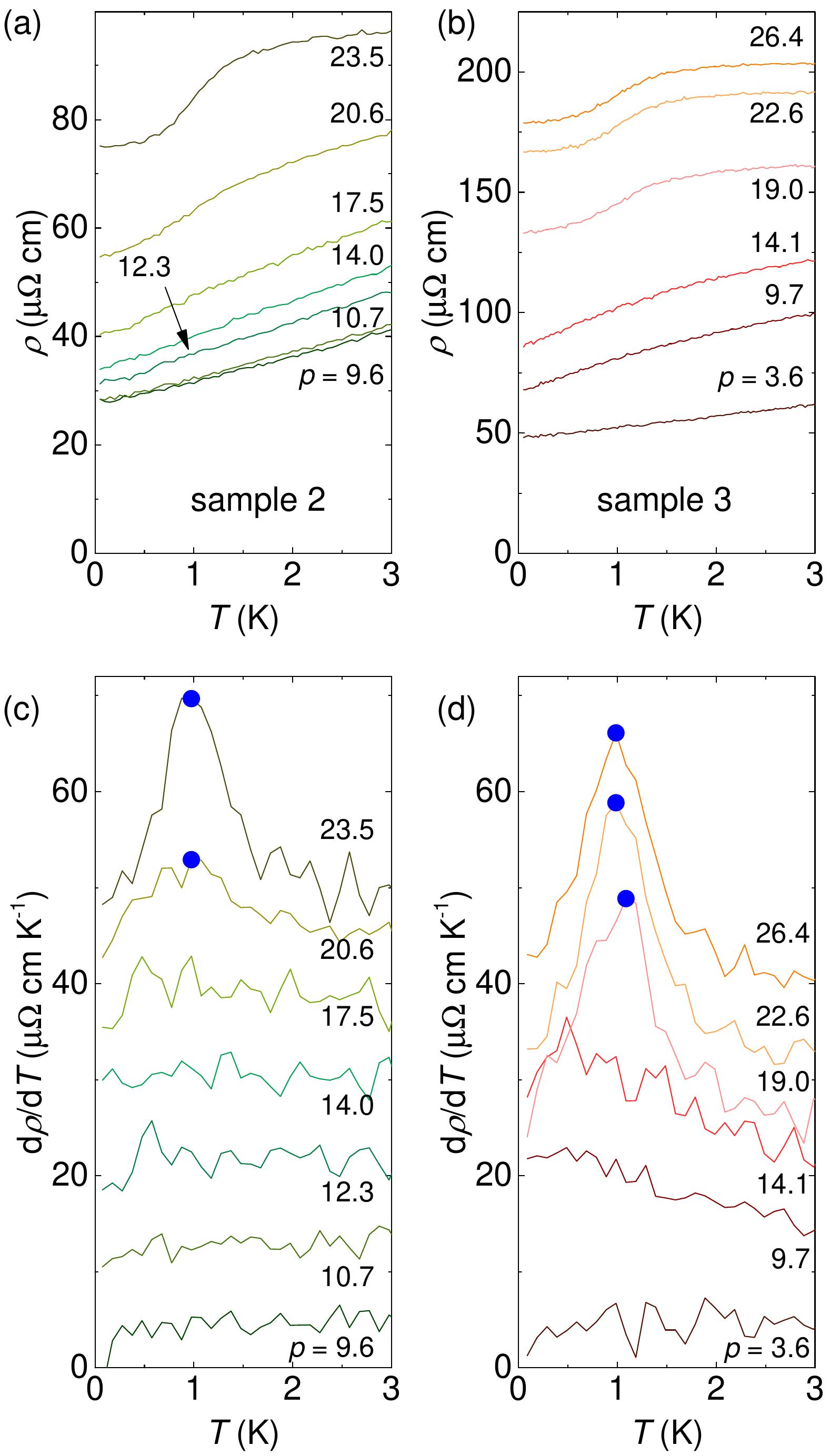}
	\end{center}
	\caption{\label{dRho_T}(Color online) Low temperature resistivity of (a) sample 2 and (b) sample 3 and (c),(d) their corresponding temperature derivatives. Solid blue circles in (c) and (d) represent the criteria (peak of the $d\rho$/$dT$) used to obtain the transition temperature. Curves in (c) and (d) are offset by increments of 8\,$\mu\Omega$\,cm\,K$^{-1}$ for clarity.} 
\end{figure}

\begin{figure}[htb]
	\begin{center}
		\includegraphics[width=8.5cm]{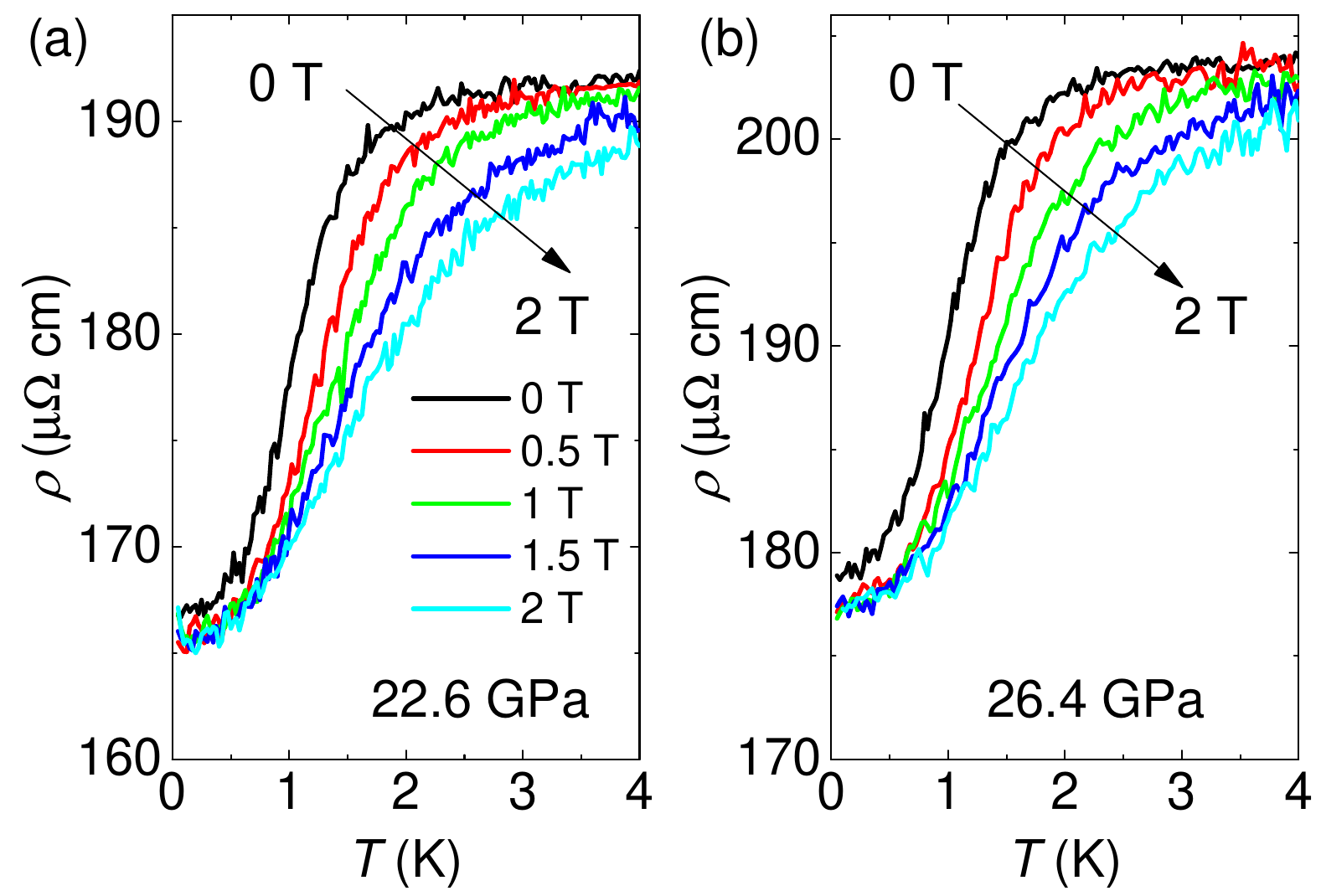}
	\end{center}
	\caption{\label{Rho_with_H}(Color online) Temperature dependence of the resistivity at various magnetic fields for (a) $p$\,=\,22.6\,GPa and (b) $p$\,=\,26.4\,GPa.} 
\end{figure}

	The evolution of the low temperature resistivity for samples 2 and 3 are shown in Figs.\,\ref{dRho_T}\,(a) and (b) respectively. For $T<T_\text{max}$ and $p<$ 18.2 GPa, the resistivity of both samples decreases with decreasing temperature and there is no pronounced anomaly down to 50\,mK. When the pressure exceeds $p_\text c=18.2\pm0.8$ GPa, the resistivity shows a kink/sharp drop, suggesting a loss of spin-disorder-scattering and magnetic ordering at \Tc. The peak in the temperature derivative of the resistivity $d\rho/dT$ is used to determine the ordering temperature, \Tc{}, as shown in Figs.\,\ref{dRho_T}\,(c) and (d). As can be seen, the peak position does not change with the pressure and remains essentially the same up to 26\,GPa. From the resistivity measurements, we cannot determine the nature of the magnetic transition, however, as shown in Fig.\,\ref{Rho_with_H}, application of magnetic field broadens the kink/sharp drop of the resistivity and moves it to  higher temperatures, which suggests that this is not a structural phase transition. Instead, this is typical behavior for a ferromagnetic transition.

\begin{figure}[htb]
	\begin{center}
		\includegraphics[width=8.5cm]{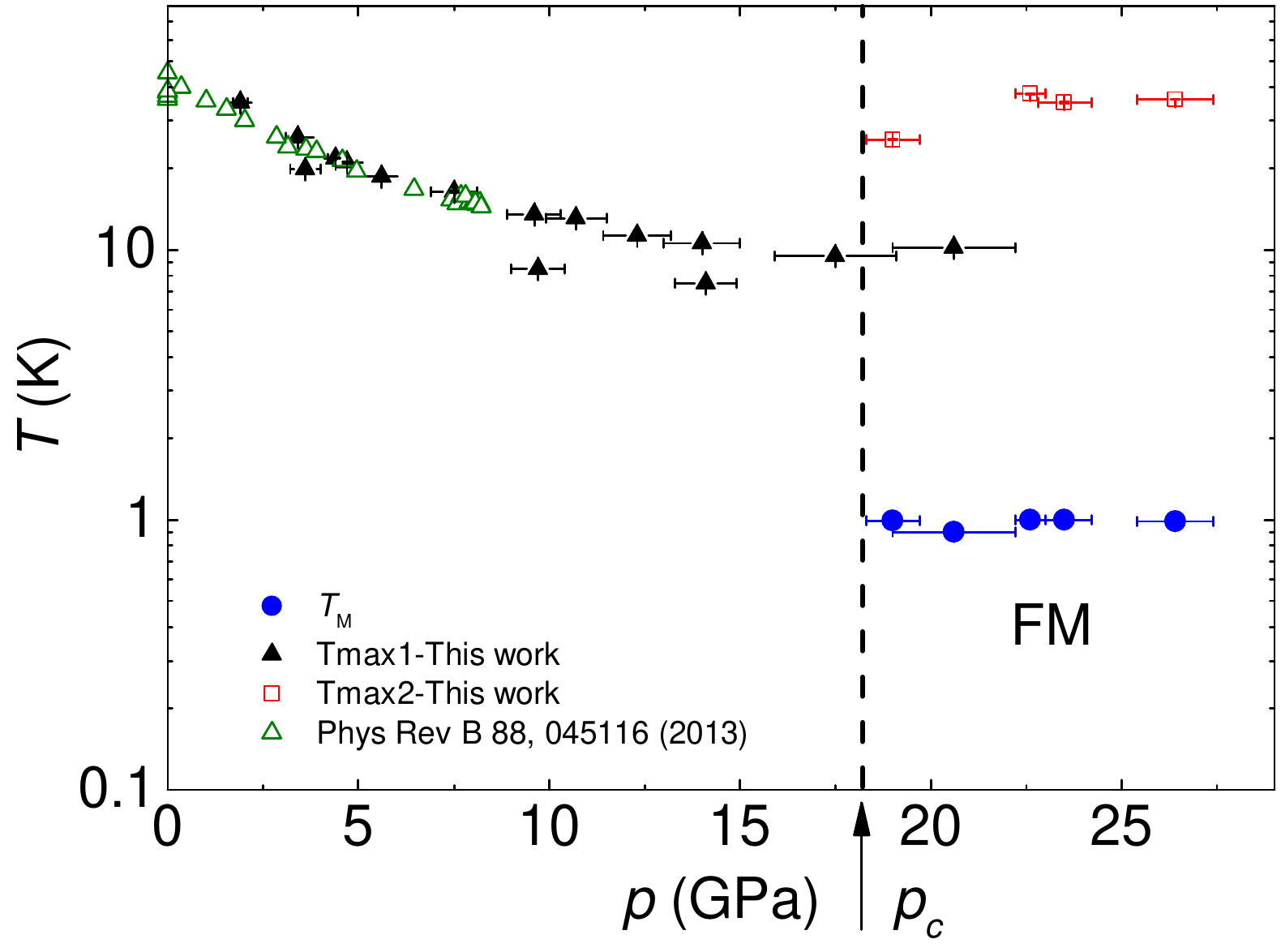}
	\end{center}
	\caption{\label{TP_phase}(Color online) Temperature-pressure phase diagram of \YbFeZn as determined from resistivity measurement. $T_\text M$, $T_\text{max1}$ and $T_\text{max2}$ are obtained using the criteria described in Fig. \ref{Rho_mag} and \ref{dRho_T}. The green open triangles are obtained from Ref.\,\onlinecite{Kim2013PRB88}. Vertical arrow represents the critical pressure \pc\,=\,18.2$\pm$\,0.8\,GPa for ferromagnetic transition as well as 4f-localization. The error bars of $p$ are determined by performing ruby fluorescence on several locations inside the sample space. The error bars of temperature are determined as half the data spacing.}
\end{figure}

A pressure-temperature phase diagram can be constructed and shown in Fig.\,\ref{TP_phase} using the data from Figs. \ref{Rho_mag} and \ref{dRho_T} as well as data from Ref. \onlinecite{Kim2013PRB88}. Black solid triangles and red open squares represent the data obtained from Fig.\,\ref{Rho_mag}. The \Tmax{} obtained from Ref.\,\onlinecite{Kim2013PRB88} is represented by open green triangles.

Figs.\,\ref{TP_phase} and \ref{R300K} demonstrate three changes that take place as pressure increases through $p$\,$\sim$\,20\,GPa. At low  temperatures ($\sim$\,1\,K) there is the sudden appearance of a transition that is arguably ferromagnetic. At intermediate temperatures there is the disappearance of an $\sim$\,10\,K resistive maximum associated with the Kondo effect and the appearance of an 30-40\,K resistive feature that is most likely associated with CEF splitting\,\cite{Cornut1972PRB,Hanzawa1985}. At higher temperatures, all the way up to room temperature, there is a marked increase in resistivity that starts around 15\,GPa and appears to saturate by $\sim$ 25\,GPa. Taking these three observations together, our results strongly suggest that by $\sim$\,20\,GPa there is a band structure change associated with the droping of the Yb-4$f$-levels below the Fermi level. As a result, the Yb-4$f$-levels stop being hybridized and the system enters the magnetic regime in the Doniach phase diagram.

These results can be put in the context of the $R$Fe$_2$Zn$_{20}$ ($R$\,=\,Gd-Tm) series which shows a clear de Gennes scaling of its ferromagnetic ordering temperature\,\cite{Jia2007Nat,Jia2009PRB}. According to the de Gennes scaling, if Yb$^{3+}$ were to be purely local-moment-like, \YbFeZn{} would order ferromagnetically at about 1\,K. This is essentially what we find for $p>p_\text c$. The suggested pressure-induced ferromagnetic ordering in YbFe$_2$Zn$_{20}$ is not too surprising, if we look at other intermetallic compounds in $R T_2$Zn$_{20}$ family\cite{Jia2008PRB,Jia2009PRB}. Taking Gd$T_2$Zn$_{20}$ series for example, ferromagnetic ordered ground state is found for members in the iron column ($T$ = Fe, Ru and Os) while antiferromagnetic ordered ground state in the cobalt column ($T$ = Co, Rh and Ir) \cite{Jia2008PRB}. Moreover, the pressure-induced ordered states in YbCo$_2$Zn$_{20}$ and YbIr$_2$Zn$_{20}$ have also been suggested as AFM ordering\cite{Taga2012JPSJ,Honda2012JPSJ}. Taking YbFe$_2$Zn$_{20}$ in this study together, $R T_2$Zn$_{20}$ family seems to follow the rule that for the iron column members, ferromagnetic ordering is expected, while for the cobalt column members, antiferromagnetic ordering is expected.

\begin{figure}[htb]
	\begin{center}
		\includegraphics[width=8.5cm]{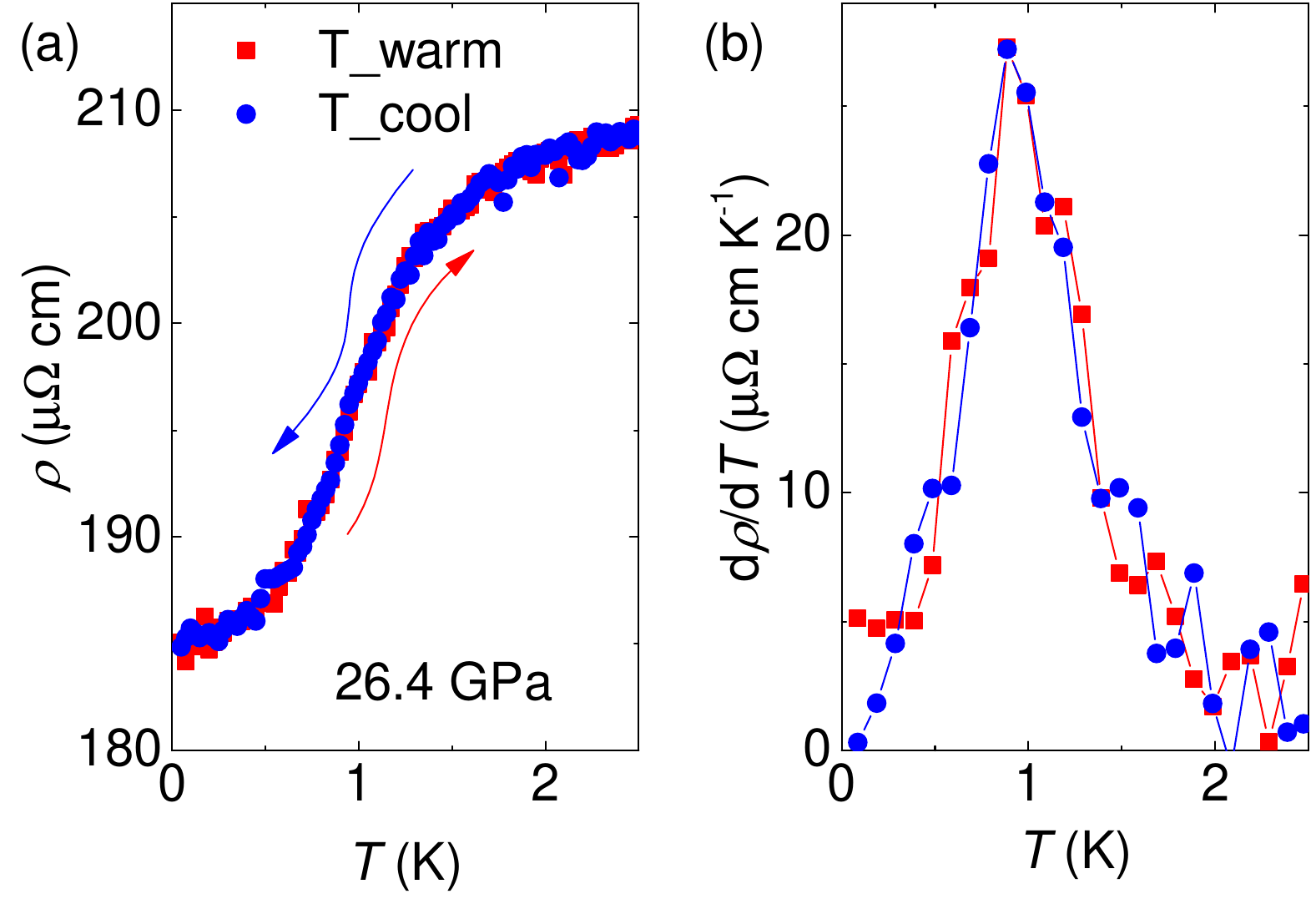}
	\end{center}
	\caption{\label{p24_hysteresis}(Color online) (a) Temperature dependence of the resistivity at 26.4\,GPa for increasing (red solid squares) and decreasing (blue solid circles) temperature. (b) Temperature derivative of $\rho$ for both increasing (red solid squares) and decreasing (blue solid circles) temperature.}
\end{figure}

\Tc{} appears to abruptly develop above \pc{} suggesting the first-order nature of the quantum phase transition at \pc. This is consistent with the growing number of examples of avoided quantum criticality in ferromagnetic metals\cite{Kaluarachchi2017NatComm,Taufour2010PRL,Kotegawa2011JPSJ,Kittler2013PRB}. According to the current theoretical understanding, a continuous  PM to FM transition is not possible at $T$\,=\,0\,K, when suppressing the FM phase with a clean parameter such as pressure\,\cite{Brando2016RMP}. Two possibilities have been proposed\,\cite{Belitz1997PRB,Belitz1999PRL}; either the transition becomes of the first order\,\cite{Taufour2010PRL,Kabeya2012JPSJ,Taufour2016PRL,Kaluarachchi2018PRB} or the modulated magnetic phase appears to replace the ferromagnetic one\,\cite{Kotegawa2013JPSJ,Kaluarachchi2017NatComm,Friedemann2017NatPhy}. In order to check for hysteresis effects (first order transition), the resistivity measurements were carried out with both increasing and decreasing temperatures at 26.4\,GPa (see Fig.\,\ref{p24_hysteresis}). However, no hysteretic behavior is observed. This could be due to a weak-first-order transition, where the hysteresis is small and may not be detected experimentally. Also, it could be due to 26.4\,GPa being higher than the pressure that corresponds to the tricritical point, so that, at 26.4 GPa, the transition is second order in temperature.

\begin{figure}[!htb]
	\begin{center}
		\includegraphics[width=8.5cm]{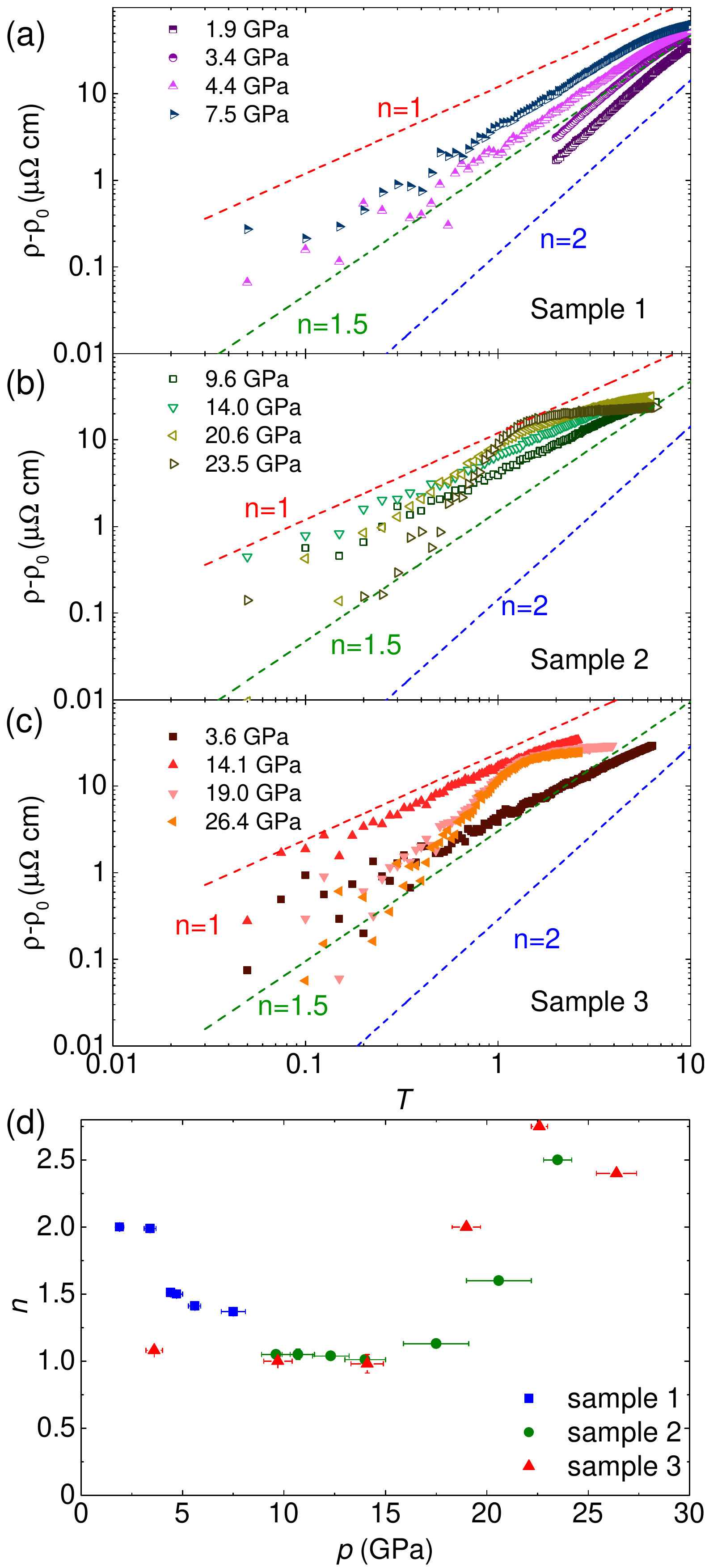}
	\end{center}
	\caption{\label{Tn}(Color online) $\rho$-$\rho_0$ vs $T$ for (a) sample 1 (b) sample 2 and (c) sample 3 in log-log scale for selected pressures. The Red, green and blue dashed lines are guides showing slopes for various  low-temperature exponents, $n$\,=\,1, 1.5 and 2 respectively. (d) Evolution of $n$ as a function of pressure. Corresponding $n$ is obtained by fitting (a)-(c) with $\rho$-$\rho_0$\,=\,$A$$T^n$. For $p>p_\text c$, fitting was done for $T<T_\text M$.}
\end{figure}

Let's consider more details about the temperature variation of the resistivity.  Figs \ref{Tn}\,(a)-(c) show the $\rho$-$\rho_0$ versus $T$ for sample 1, 2 and 3 to emphasize the low-temperature exponent, $n$, which appears as the slope on a log-log scale. At low-pressures (in Fig.\,\ref{Tn}\,(a) $p<$\,3.4\,GPa), $\rho$-$\rho_0$ obeys $T^2$ and for the intermediate pressures ($p\sim$\,6\,GPa) it follows $T^{1.5}$. For higher pressures  (\,9\,GPa $<p<$\,18\,GPa), $\rho$-$\rho_0$ shows linear $T$ dependence over a wide range of temperature. A $T$ -linear resistance has been observed in several compounds, such as; CeCoIn$_5$\,\cite{Sidorov2002PRL}, CeRhIn$_5$\,\cite{Knebel2008JPSJ,Park2008Nature}, YbRh$_2$Si$_2$\,\cite{Trovarelli2000PRL,Gegenwart2002PRL,Custers2010PRL}, YbAgGe\cite{Budko2004PRB,Niklowitz2006} and  CeNi$_2$Ge$_2$\,\cite{Grosche2000}. Evolution of the temperature power-law exponent $n$ with pressure is summarized in Fig.\,\ref{Tn}\,(d).  The value of $n$ is obtained from a sliding window fit to $\rho$-$\rho_0$\,=\,$A$\,$T^n$, where $\rho_0$ is obtained from the fit at the lowest temperature. Since the data have been taken down to 0.05\,K, the value of $\rho_0$ can be obtained more accurately than was possible for Ref. \onlinecite{Kim2013PRB88}. As can be seen, $n$ is clearly deviating from 2 for higher pressures. Power-law analysis from Ref.\,\onlinecite{Kim2013PRB88} indicates $n$\,=\,2 even at $p$\,$\sim$\,8\,GPa (see Figs.\,5 and 9 in Ref.\,\onlinecite{Kim2013PRB88}). This most likely led to the low estimated value of \pc{} based on the divergence of $A$-coefficient in Ref. \onlinecite{Kim2013PRB88}. %Apart from this, inhomogeneous pressure distribution due to the solid pressure medium in our experiment could also responsible for these discrepancies. 
For $p>p_\text c$ the low temperature loss of spin disorder feature has $\propto T^n$ behavior most likely associated with spin excitation scattering not too far below $T_\text c$. As indicated in Fig. \ref{Tn} (d), as pressure is increased above $p_\text c$, $n$ quickly deviates from 1.
 
%The unusual linear $T$ dependence of $\rho$($T$) observed close to the ferromagnetic instability is not expected in the spin-fluctuation theory\,\cite{Moriya1985,Moriya1995JPSJ}. Theoretically, a $T^{5/3}$  dependence is predicted for $\rho(T)$ in the quantum critical regime for a 3D ferromagnet\,\cite{Mathon1968,Moriya2003}.  There are several non-Fermi liquid (NFL) theories which have different microscopic origins and can describe the behavior of many of these systems with reasonable success. For example, models involving a quantum phase transition\,\cite{Hertz1976,Millis1993PRB}, the Kondo disorder model\,\cite{Miranda1996,Miranda1997PRL} or Griffiths' phase model\,\cite{CastroNeto1998PRL} have been proposed to describe the low-temperature NFL behavior. Also, it has become clear that the influence of disorder often leads to unconventional behavior\,\cite{Rosch1999PRL}. Therefore, we would like to point out that the experimental and theoretical arguments regarding the observed anomalies related to the existence of $n$\,=\,1 should be considered with proper caution. 
	
From the constructed $p-T$ phase diagram (Fig. \ref{TP_phase}), it is shown that for YbFe$_2$Zn$_{20}$, at the low temperature region, the associated Kondo temperature $T_\text {max1}$ is first suppressed with increasing pressure. At $p_\text c$, a possibly ferromagnetic transition $T_\text M$ suddenly appears at $\sim$ 1 K and stays unchanged with further increasing pressure. This suggests that for YbFe$_2$Zn$_{20}$, the quantum criticality is avoided by going through a first-order QPT under pressure, which is in contrast to the YbCo$_2$Zn$_{20}$ and YbIr$_2$Zn$_{20}$, where they enter AFM ordered states through QCP\cite{Saiga2008JPSJ,Taga2012JPSJ,Honda2012JPSJ}. At high temperature, a continuously increase of the resistivity with pressure was observed for $p\gtrsim p_\text c$ (Fig. \ref{R300K}), suggesting that the suppressing of hybridization and developing of the Yb$^{3+}$ local moment is more continuous in nature.

\section{Conclusions}
In summary, we have measured the resistivity of \YbFeZn{} up to $\sim$\,26\,GPa and down to 50\,mK. Above a critical pressure, \pc\,=\,18.2$\pm$\,0.8\,GPa, we observed the resistivity anomaly at \Tc\,$\sim$\,1\,K, which remains constant with increasing pressure. This anomaly appears to correspond to a ferromagnetic transition, since the application of magnetic field broadened the transition and moved it to higher temperature. Increasing pressure drives the \Tmax{}, the associated Kondo temperature, to lower values and flattening at pressures up to \pc{} indicating a decrease of the hybridization strength. Above \pc{}, \Tmax{} abruptly increases with pressure. In this pressure range $T_\text{max}$ can be attributed to the crystal electric field effects. In heavy fermion non-magnetic phase, the low temperature power law exponent is deviated from the Fermi liquid behavior for $p$\,$>$\,3.4\,GPa and reached $n$\,=\,1 for  9\,GPa\,$<$\,$p$\,$<$\,\pc{}. The reason for this unusual exponent value, $n$\,=\,1, over large range of pressure is not clear so far. Additionally, our data suggests that at $\sim p_c$ there is a band structure change associated with the dropping of 4f-levels below the Fermi level.

\section*{ACKNOWLEDGMENTS}
We would like to thank A. Kaminski for assisting in preparation the gaskets that are used for the pressure cell. This work was carried out at the Iowa State University and supported by the US DOE, Basic Energy Sciences, Materials Science and Engineering Division under contract No. DE-AC02-07CH11358. L. X. was supported, in part, by the W. M. Keck Foundation. The research at the Geophysical Laboratory  was supported by DOE/BES under contract No. DE-FG02-99ER45775. A. G. G. acknowledges support of RSF 16-12-10464 grant. For preparation of high-pressure cells the  facilities  of  Center for Collective Use " Accelerator Center for Neutron Research of the Structure of Substance and Nuclear Medicine" of the INR RAS were used.

%\newpage	
%\pagebreak
\clearpage

\appendix
\renewcommand\thefigure{A.\arabic{figure}}    
\section*{\\Appendix}
\setcounter{figure}{0}

\label{Appendix}
	\begin{figure}
		\begin{center}
			\includegraphics[width=8.5cm]{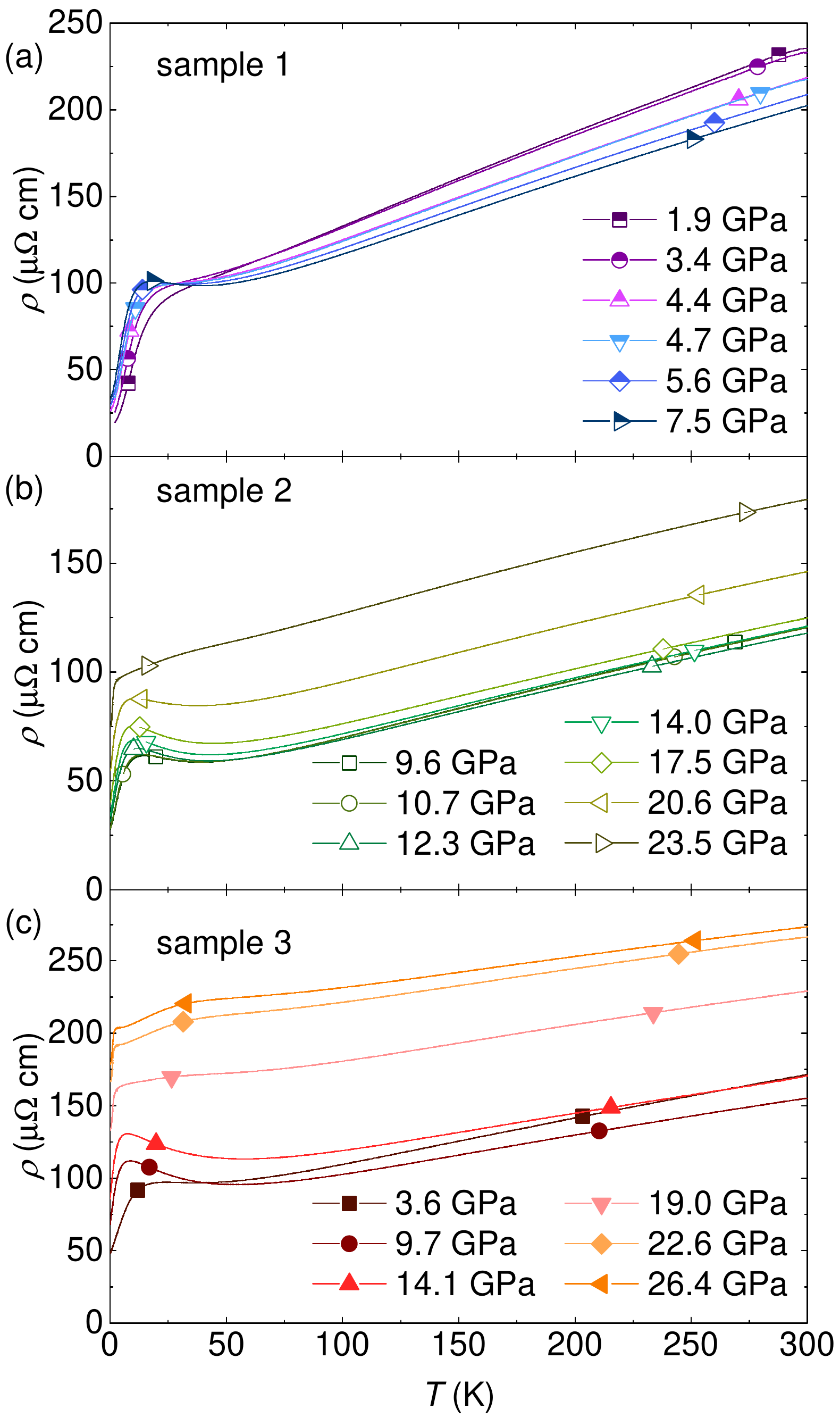}
		\end{center}
		\caption{\label{Rho_T_UN}(Color online) Temperature dependence of the resistivity of three different samples of \YbFeZn: (a) sample 1, (b) sample 2 and (c) sample 3.} 
	\end{figure}
	
	\begin{figure}
		\begin{center}
			\includegraphics[width=8.5cm]{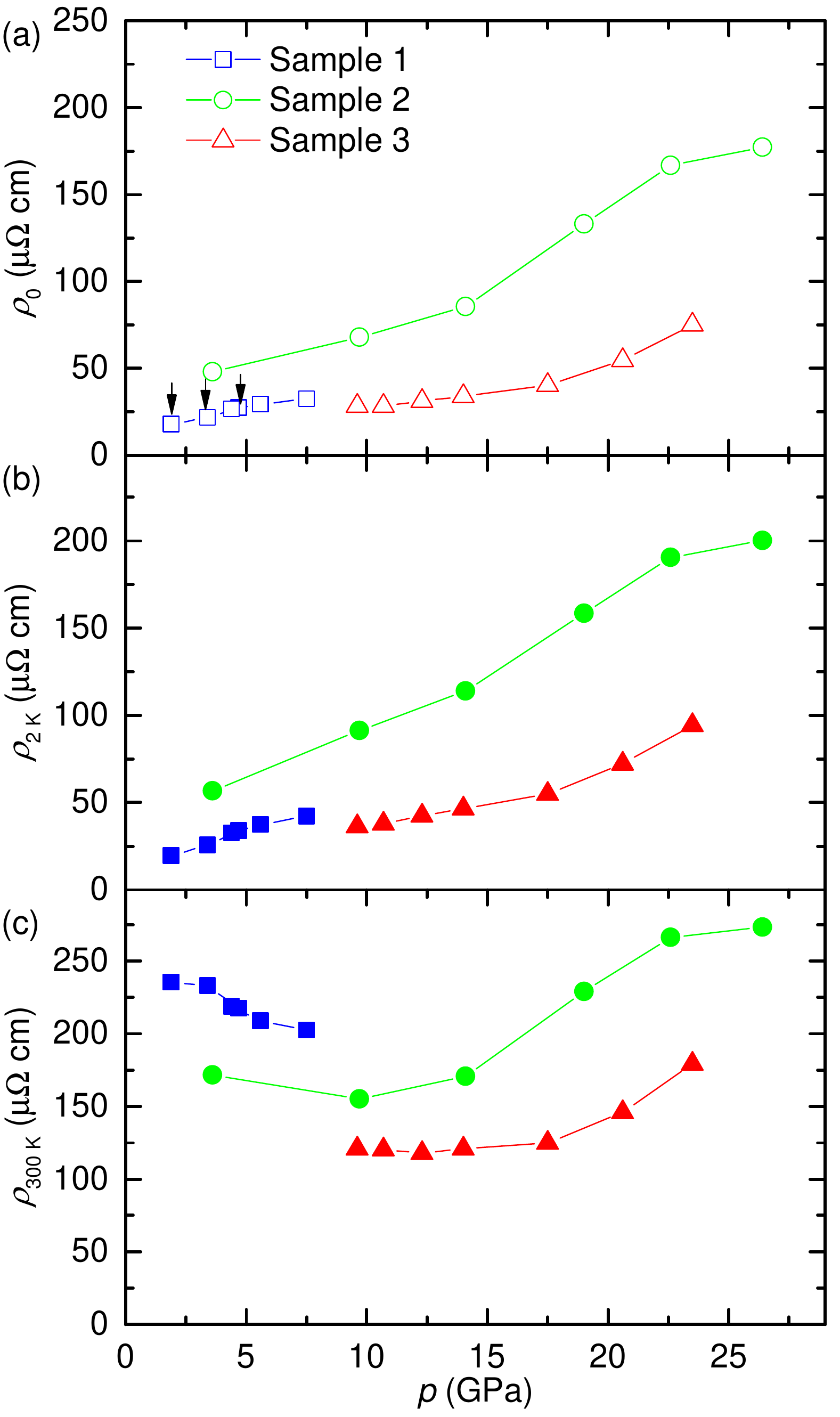}
		\end{center}
		\caption{\label{RRR}(Color online) Pressure dependence of resistivity $\rho$ at different fixed timperatures. (a) $\rho_0$ values were obtained by extrapolating low temperature $\rho(T)$ data to 0 K. For sample 1, the first three pressures, measurements were done down to 2 K (arrows in (a)), for the rest, measurements were done down to 0.05 K. (b)Resistivity $\rho$ at 2 K. (c)Resistivity $\rho$ at 300 K.}
	\end{figure}
Fig. \ref{Rho_T_UN} presents the temperature dependence of the resistivity of three different samples without normalization. Fig. \ref{RRR} presents the pressure dependence of the resistivity values at different fixed temperatures. (a) Extrapolated 0 K resistivity values $\rho_0$ by fitting low temperature $\rho(T)$ data, (b) $\rho$ at 2 K, (c) $\rho$ at 300 K. Fig. \ref{R300K} is the normalized version of Fig. \ref{RRR} (c) obtained by dividing the 300 K resistivity value at the lowest pressure for each sample.

    %\begin{figure}
	%	\begin{center}
	%		\includegraphics[width=8.5cm]{RRR}
	%	\end{center}
	%	\caption{\label{RRR}(Color online) Pressure dependence of $\rho$ at (a) 0\,K, (b) 2\,K and (c) 300\,K. The $\rho_0$ ($\rho$ at 0\,K) is obtained by low temperature fitting. This value can be obtained more accurately since most of the data have been taken down 0.05\,K. For sample 1, for first three pressures, resistivity measured down to 2\,K (vertical arrows in (a)).} 
	%\end{figure}
	
%Fig. \ref{RRR} represents the resistivity at three different fixed temperatures: (a) $\rho$\,=\,0\,K, (b) $\rho$\,=\,2\,K and (c) $\rho$\,=\,300\,K. Fig. \ref{R300K} is the normalized version of Fig.\,\ref{RRR}\,(c). 
\bibliographystyle{apsrev4-1}

%\bibliography{C:/Users/uskalu/BoxSync/MyRef/MyDataBase}    %C:\Users\uskalu\Box Sync\MyRef
%\bibliography{output}
%

\end{document}